\documentclass[onecolumn]{revtex4}
\usepackage{amssymb,amsmath}
\usepackage{graphicx}

\begin{document}


\title{Runaway evaporation for optically dressed atoms}

\author{David Wilkowski}

\address{Institut Non Lin\'eaire de Nice, Universit\'e de Nice Sophia-Antipolis, CNRS, Valbonne 06560, France.}
\address{Centre for Quantum Technologies, National University of Singapore , Singapore 117543, Singapore.}

\begin{abstract}
Forced evaporative cooling in a far-off-resonance optical dipole trap is proved to be an efficient method to produce fermionic- or bosonic-degenerated gases. However in most of the experiences, the reduction of the potential height occurs with a diminution of the collision elastic rate. Taking advantage of a long-living excited state, like in two-electron atoms, I propose a new scheme, based on an optical knife, where the forced evaporation can be driven independently of the trap confinement. In this context, the runaway regime might be achieved leading to a substantial improvement of the cooling efficiency. The comparison with the different methods for forced evaporation is discussed in the presence or not of three-body recombination losses.
\end{abstract}

\maketitle


Quantum degenerate gases are now routinely produced in tens of laboratories across the world. The current methods, even if they differ in their approaches, rely, on their final stage for forced evaporative cooling in a conservative trap. Two classes of traps; a high-field-seeking magnetic trap and a far-off-resonance optical dipole trap, are used independently, or in combination. In addition to design constraints, such as optical access, robustness and reproducibility, the experimental apparatus should produce a degenerate gas with the largest number of atoms and the best duty cycle. The new evaporation scheme, I describe in this paper, goes in that direction.

Forced evaporative cooling in magnetic trap is usually performed using a RF probe or \emph{knife} to coupled a trapping state to an anti-trapping one. The frequency of the RF \emph{knife} is swept to lower the energy barrier of the truncated potential. During forced evaporation the confinement of the ultracold gas remains almost unchanged. In an optimized scenario, when the evaporation occurs on the whole surface where the RF resonance is fulfilled,  the spatial density increases faster than the inverse of the mean velocity leading to a net increase of the elastic collision; the so-called runaway regime \cite{PhysRevA.53.381}.

In magnetic traps, the atomic magnetic moment is frozen to fulfill the trapping condition. Spin degrees of freedom can be released in far off-resonance optical dipole trap. In this case, the interplay between many-body systems and atomic internal degrees of freedom open the door of a rich variety of new phenomena in ultracold atoms physic. In particular, this approach has been used to produce spinor condensates \cite{stenger1999spin,PhysRevLett.92.140403,PhysRevLett.92.040402}, fermionic spin mixtures and molecular ultracold samples \cite{PhysRevLett.88.120405, PhysRevLett.91.080406, PhysRevLett.91.240401, greiner2003emergence, jochim2003bose, PhysRevLett.97.180404, PhysRevLett.97.120402, PhysRevLett.102.020405}.

For some atomic species, evaporation toward quantum degeneracy is difficult in a magnetic trap. In those cases, evaporation in an optical dipole trap is an alternative. Elements, like Cs \cite{PhysRevLett.80.1869,weber2003bose} or $^{85}$Rb \cite{PhysRevLett.85.1795}, for example, have unfavorable collision properties and their scattering length needs to be tuned thanks to a magnetically induced Feshbach resonance \cite{chin2008feshbach}. Two-electron atoms, which carry any electronic magnetic momentum in their fundamental level, have also to be cooled with all optical methods. Recently, BEC of $^{40}$Ca \cite{kraft2009bose}, $^{84}$Sr \cite{PhysRevLett.103.200401,MartinezdeEscobar}, $^{174}$Yb \cite{PhysRevLett.91.040404} and $^{170}$Yb \cite{PhysRevA.76.051604}, Fermi sea of $^{87}$Sr \cite{desalvo2010degenerate} and $^{173}$Yb \cite{PhysRevLett.98.030401} and degenerate mixtures of $^{88}$Sr-$^{87}$Sr \cite{mickelson2010bose}, $^{176}$Yb-$^{174}$Yb, $^{173}$Yb-$^{174}$Yb \cite{PhysRevA.79.021601} have been reported.

In dipole traps no energy selective coupling to an anti-trapping state, similar to the RF \emph{knife}, has been implemented so far. Evaporation takes place because hottest atoms, with a mechanical energy higher than the potential height, are escaping the trap. Thus, the dipole trap might fulfil at least two major requirements. First, the potential height has to be in the order of few thermal energy, $k_BT$ such that the evaporation rate is significant. Second, the spatial confinement and the compression of the atomic gas have to be such that elastic collisions can take place leading to fast thermalization. Different strategies have been used to improve the preparation of the cold sample before evaporation: For alkali atoms, Raman \cite{weber2003bose} or polarization gradient cooling \cite{PhysRevA.71.011602} have been implemented to lower the temperature. For two-electron atoms, low temperatures are achieved with Doppler cooling on the intercombination lines \cite{kraft2009bose,PhysRevLett.103.200401,MartinezdeEscobar,PhysRevLett.91.040404,katori1999magneto,loftus2004narrow,chaneliere2008three}. Improvement of the spatial density in the dipole trap is also reported using adiabatic compression by dynamically changing the trap geometry, for example with an additional dimple trap \cite{weber2003bose} or a Zoom lens to displace one of the beam waist position \cite{PhysRevA.71.011602}. The forced evaporation is performed by dynamically lowering the potential height.

In the original experimental realizations of the BEC \cite{PhysRevLett.87.010404} and the Fermi gas \cite{PhysRevLett.88.120405} in all optical devices, the gas was confined in a cross dipole trap to insure a good 3D confinement. The forced evaporation was carried out by lowering the power $P$ of the trap light field. As a consequence the trapping frequencies,
\begin{equation}
\label{eq_wP}
\omega\propto\sqrt{P},
\end{equation}
and the trap confinement are also reduced. In this situation and in sharp contrast with RF \emph{knife}, the elastic collision rate decreases during the forced evaporation and no runway occurs \cite{PhysRevA.64.051403}. Thus, it is crucial to counteract the reduction of efficiency of cooling during evaporation with a very good starting elastic collision rate.

New optical based trap schemes have been successfully implemented to limit and event almost suppress the reduction of trapping confinement during the forced evaporation. Those realizations have the same underlying idea which consists in decoupling the potential height driving the evaporation, and the trap confinement controlling the thermalization rate. In \cite{PhysRevA.78.011604} a tilt trap was used where spin polarized atoms of Cesium are held in a fixed dipole trap with a superimposed varying magnetic field gradient. The extra constant force pulls the atoms out of the trap. In all optical schemes, different versions of the dimple trap have been explored. In those schemes, the trap is made of at least two independent laser beams: a tightly confining one for 2D trapping and a wider one to close the trap in the third dimension. In an ideal scenario, only the power of wider beam is reduced \cite{kraft2009bose,PhysRevLett.91.040404}. Cl\'ement and co-authors \cite{PhysRevA.79.061406} report 3D confinement and runaway evaporation in a stubble misaligned version of the dimple trap which works in a similar manner as the tilt trap of \cite{PhysRevA.78.011604}.

In this paper, I propose a new method for runaway evaporation in an optical trap. In contrast with the methods implemented so far successfully, evaporation is not based on lowering the trapping lasers power but on a potential truncated with an optical \emph{knife}. The paper is organized as follow: In section \ref{Sec_Dipole_Knife}, I define the general requirements to implement the forced evaporation with an optical \emph{knife} in a dipole trap. On the basis of a thermodynamical quasi equilibrated state, I derive rate equations which govern the evolution of some macroscopic quantities of the ultracold gas in the trap (section \ref{Sec_Evap_fix_Vary}). Different evaporation strategies can be put in place according to the strength of the three-body recombination loss rate. Finally, I will draw the conclusion of this work in section \ref{Sec_Conclusion}.

\section{Dipole trap with optical \emph{knife}}\label{Sec_Dipole_Knife}

I consider a two levels system in the optical domain where the excited (ground) state, labeled $|e\rangle$ ($|g\rangle$), has a radiative lifetime long enough to disregard any spontaneous emission. Like in other evaporation schemes, far off-resonance lasers insure the 3D confinement of the cold gas in the $|g\rangle$ state. It is crucial however that $|e\rangle$ has to be an anti trapping state. The $|g\rangle\rightarrow|e\rangle$ transition is driven with a quasi resonant optical field at a detuning $\delta$ with respect to the bare frequency difference of the two states. In the RWA approximation, the eigenenergies are
\begin{equation}
E_\pm=\frac{s}{2}\left(1\pm\sqrt{\left(1-4\frac{p-\frac{\Omega^2}{4}}{s^2}\right)}\right),
\end{equation} where $s=V_e(\vec{r})-\delta+V_g(\vec{r})$ and $p=(V_e(\vec{r})-\delta)\cdot V_g(\vec{r})$ are respectively the sum and the product of the ground and excited state energies dressed with the far off-resonance trapping lasers. $\Omega$ is the Rabi frequency of the quasi-resonant field. This field is supposed to give any extra light shift contribution. In the adiabatic regime, an atom, initially in the trap, \emph{i.e.} in the $|g\rangle$ state, is resonantly brought into the anti trapping state $|e\rangle$ at a distance where the avoid crossing occurs ($V_g(\vec{r})=-V_e(\vec{r})+\delta$). Hence forced evaporation might be achieved by sweeping the detuning of the quasi-resonant field from zero toward a blue value.

\begin{figure}
\begin{center}
\includegraphics[scale=1]{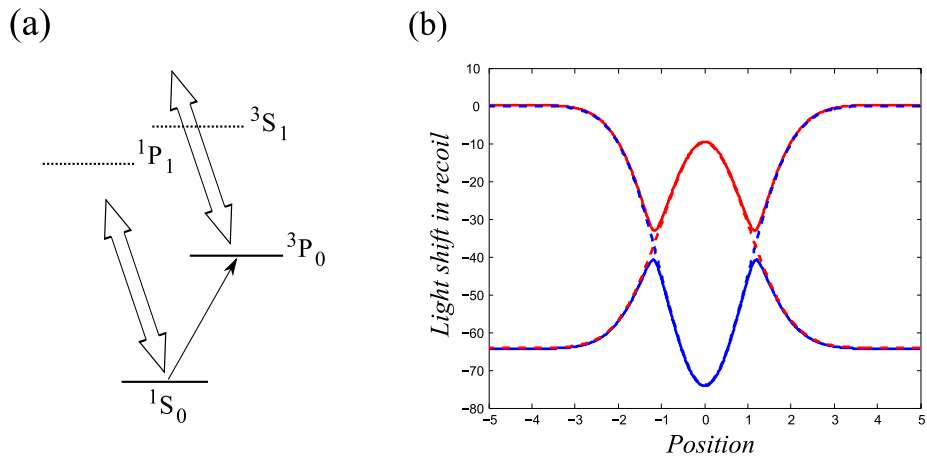}
\caption{(a) Simplified energy diagram and states involved in forced evaporation of $^{87}\textrm{Sr}$ with an optical \emph{knife} in quasi resonance with the $^1\!S_0\rightarrow\,^3\!P_0$ transition. The coupling laser has a power of $50\textrm{mW}$ focus on a $50\mu\textrm{m}$ spot size. The Rabi frequency is $\Omega=2\pi\!\cdot\! 40\textrm{kHz}$. The dipole trap is produced with $660\textrm{nm}$ crossed laser beams focus on a $50\mu\textrm{m}$ spot size. The total power is $1\textrm{W}$ for a mean radial trap frequency of $270\textrm{Hz}$. (b) shows the energy radial distribution, in waist unit, of the potential in energy recoil unit. The dash blue (red) curve corresponds to the $^1\!S_0$ ground ($^3\!P_0$ excited) state in interaction with the trapping laser only. The plain curves indicate the adiabatic potential dressed with the optical \emph{knife}. In this example, the detuning of the \emph{knife} is $\delta=2\pi\cdot 340\textrm{kHz}$.}
\label{87Sr}
\end{center}
\end{figure}

For a practical implementation, two-electron atoms are a straightforward choice since one of the long living states in the triplet spectrum can be used as an excited state. The $^3\!P_0$ state is in particular interest because of the absence of an electronic magnetic momentum. Thus the forbidden $^1\!S_0\rightarrow\,^3\!P_0$ transition can be almost decoupled to stray magnetic fields. For the odd isotopes, the $^1\!S_0\rightarrow\,^3\!P_0$ transition, due to hyperfine mixing, becomes weakly allowed to single photon direct excitation. The bare linewidth of the transition is in the range of few mHz \cite{PhysRevA.69.042506} which prevents any spontaneous emission during evaporation and provides enough coupling strength among the two states. For the even isotopes, the $^1\!S_0\rightarrow\,^3\!P_0$ transition is strongly forbidden. In that case, the $^3\!P_0$ state needed to be mixed to another short living states as depicted in \cite{PhysRevLett.96.083001,PhysRevLett.94.173002}.

Beside the issues of getting favorable conditions for evaporation, one has also to keep in mind that the light shift due to the off-resonance lasers should create an anti-trapping state for $|e\rangle$. Supposing that $|g\rangle$ is a trapping state, this condition automatically gets fulfilled for a two-level system where as in real atoms light shifts values can be engineered almost at will according to the presence of the other excited levels. If far infrared traps have to be excluded because they give the same negative sign of the electrical polarizability for both states at play, then it seems, however, that this new constraint can be overcome in most cases. For Yb, a popular near infrared laser is suitable \cite{Gerbier2010Gauge} where as for Sr a dipole trap made with a red laser might be an alternative. For the latter case, the $5s^2\,^1\!S_0\rightarrow 5s6p\,^1\!P_1$ and $5s5p\,^3\!P_0\rightarrow 5s6s\,^3\!S_1$ transitions respectively at $461\textrm{nm}$ and $679\textrm{nm}$ give the major contribution to the states light shifts (see figure \ref{87Sr}a). An example of the potential, dressed with a quasi resonant \emph{knife} laser at a detuning of $\delta=2\pi\cdot 340\textrm{kHz}$, is depicted on figure \ref{87Sr}b. More values of the parameters used in this example are given in the caption figure \ref{87Sr}. I emphasize however that $\Omega=2\pi\cdot 40\textrm{kHz}$ yields an energy splitting of the two adiabatic potential curve in the microKelvin range at the avoid crossing. It is important to get a substantial energy splitting because evaporation only occurs if the escaping atom stays in the adiabatic state. Landau-Zener tunneling rate to the neighboring state can be evaluated following reference \cite{Yi2008State} and it turns out that it can be disregarded as far as $\hbar\Omega\gtrsim k_bT$. $\hbar\Omega$ has to be interpreted as the energy splitting of the two adiabatic states at resonance and $k_bT$ as the mean energy of an escaping atom \cite{PhysRevA.53.381}.

\begin{table}
\begin{center}
\begin{tabular}[t]{c c c c}
\hline
\hline
    &exact
        &approximated
            &validity\\
\hline
$\gamma_v/\gamma$
    &$\left(\eta-4\frac{P(4,\eta)}{P(3,\eta)}\right)\textrm{e}^{-\eta}$
        &$(\eta-4)\textrm{e}^{-\eta}$
            &$\eta\geq 4$\\
\hline
$\tilde{\kappa}$
    &$1-\frac{X_{ev}}{V_{ev}}$
        &$1-\frac{1}{\eta-4}\left(1-\frac{\eta^4}{24}\textrm{e}^{-\eta}\right)$
            &$\eta\geq 7$\\
\hline
$\tilde{c}$
    &$3\frac{P(4,\eta)}{P(3,\eta)}$
        &$3-\frac{\eta^3}{2}\textrm{e}^{-\eta}$
            &$\eta\geq 6$\\
\hline
$\alpha$
    &$\frac{\eta+\tilde{\kappa}}{\tilde{c}}-1$
        &$\frac{\eta+1}{3}-1$
            &$\eta\geq 9$\\
\hline
\hline

\end {tabular}
\label{table_fct}
\caption{Important parameters used in the thermodynamical model of evaporative cooling. Their expressions are given for a 3D harmonic trap. The exact expressions, in the second column, are extracted from reference \cite{PhysRevA.53.381}. The third column gives approximated expressions and last column shows their validity domain within an error smaller than $5\%$.}
\end{center}
\end{table}

\section{Evaporation in a fix or varying trap}\label{Sec_Evap_fix_Vary}

I have shown that an ultra-narrow (or clock) transition may be used as an optical \emph{knife} for evaporation in a Dipole trap. The initial high confinement is then kept almost fixed, in contrast with the original evaporation techniques in which the trap height is reduced. Let us now compare the two methods, \emph{i.e.} when the trap confinement is varying or remains fixed. For that purpose, I will establish some relations among quantities that characterize the efficiency of the evaporation. The starting point would be to set a kinetic equation of the dynamics of the ultracold gas in the finite depth trap for both cases. As far as the elastic collision rate insures a good thermalization of the gas during the evaporation, it has been showed that the distribution function in energy of the gas is well represented by a Boltzmann distribution truncated at an energy $\epsilon_t$ corresponding to the energy threshold at which the evaporation occurs \cite{PhysRevA.53.381}. Furthermore, a simple and efficient cooling ramp consists in keeping the parameter
\begin{equation}
\eta=\epsilon_t/k_bT
\label{eq_def_eta}
\end{equation}
large and constant. Here, $T$ is the temperature of the gas. Hereafter these conditions are supposed to be always fulfilled. Following references \cite{PhysRevA.53.381} and \cite{PhysRevA.64.051403}, a thermodynamical model of evaporation can be resumed in a set of rate equations,
\begin{equation}
\dot{N}=-\left[\gamma_v(n,T)+\gamma_l(n,T)\right]N\\
\label{eq_N}
\end{equation}
\begin{equation}
\dot{T}=-\left[\alpha\gamma_v(n,T)-\left(1-\frac{2}{\tilde{c}}\right)\gamma_l(n,T)\right]T,
\label{eq_T}
\end{equation}
governing the temporal evolutions of the number of atoms $N$ in the trap and their temperature. Here, $\gamma_v(n,T)$ is the evaporation rate related to the elastic collision rate $\gamma(n,T)=An\overline{v}\sigma$(see table \ref{table_fct}). $A$ is a dimensionless parameter. Its value depends on the type -fermionic or bosonic- of the atom and on the number $n_s$ of spin states involve in the s-wave scattering process. For spin polarized bosons, $A=1$. For unpolarized Fermions $A=\frac{n_s-1}{n_s}$. $\overline{v}$ and $\sigma$ are respectively the thermal velocity and the elastic scattering cross section. The expressions of $\alpha$ and $\tilde{c}$ are also given in table \ref{table_fct}. $\gamma_l(n,T)$ stands for the atoms loss rate of the three-body recombination. Other loss processes such as one-body and two-body losses, are neglected \cite{yan2009numerical}. The extra heating rate of the three-body recombination corresponds to the second term on the right hand side in equation (\ref{eq_T}). It comes from the fact that the recombination rate scales like $n^3$ and thus occurs most likely in the high density region of the cloud where the potential energy is lower than its mean value \cite{PhysRevLett.91.123201}.


\begin{table}
\begin{center}
{\normalsize \begin{tabular}[t]{@{}c c c@{}*{2}{@{}c@{}} @{}c c@{} *{2}{@{}c@{}} @{}c c c c}
\hline
\hline

            & Type
                & $N_i$
                     & $T_i$
                        & $n_i$
                            &$\Gamma_{el}^i$
                                & $N_f$
                                    & $T_f$
                                        & $n_f$
                                            & $\Gamma_{el}^f$
                                                & $t_{evap}$
                                                    & $\eta$
                                                        & Ref\\

            &
                & \footnotesize{($10^5$)}
                     & \footnotesize{($\mu\textrm{K}$)}
                        & \footnotesize{($10^{14}\textrm{cm}^{-3}$)}
                            & \footnotesize{($10^3\textrm{s}^{-1}$)}
                                & \footnotesize{($10^5$)}
                                    & \footnotesize{($\mu\textrm{K}$)}
                                        & \footnotesize{($10^{14}\textrm{cm}^{-3}$)}
                                            & \footnotesize{($10^3\textrm{s}^{-1}$)}
                                                & \footnotesize{(s)}
                                                    &
                                                        & \footnotesize{(Experimental)}\\
\hline
            & Experimental
                & 50
                    & 20
                        & 0.3
                            & \emph{n.a.}
                                & 0.09
                                    & 0.26
                                        & \emph{n.a.}
                                            & \emph{n.a.}
                                                & 1.5
                                                    & 5
                                                        & \cite{kraft2009bose}\\
        $^{40}\textrm{Ca}$
            & Varying
                & 50
                     & 20
                        & 0.3
                            & 40
                                & 0.03
                                    & 0.005
                                        & $3\cdot 10^{-4}$
                                            & $4\cdot 10^{-4}$
                                                & 67
                                                    & 5
                                                        & \\

            & Fixed
                & 50
                     & 20
                        & 0.3
                            & 40
                                & 6.5
                                    & 2
                                        & 1.4
                                            & 56
                                                & 0.002
                                                    & 5
                                                        & \\
\hline

            & Experimental
                & 10
                     & 10
                        & 1.2
                            & 3.5
                                & 3
                                    & 0.4
                                        & $\approx n_i$
                                            & 0.7
                                                & 6
                                                    & 10
                                                        & \cite{PhysRevLett.103.200401}\\
        $^{84}\textrm{Sr}$
            & Varying
                & 10
                     & 10
                        & 1.2
                            & 5.5
                                & 3
                                    & 0.4
                                        & 0.3
                                            & 0.3
                                                & 4
                                                    & 10
                                                        & \\

            & Fixed
                & 10
                    & 10
                        & 1.2
                            & 5.5
                                & 6
                                    & 2.5
                                        & 5
                                            & 14
                                                & 0.2
                                                    & 10
                                                        & \\
\hline
            & Experimental
                & 30
                    & 5
                        & 0.4
                            & 1
                                & 3
                                    & 0.4
                                        & \emph{n.a.}
                                            & \emph{n.a.}
                                                & 3
                                                    & 7.5
                                                        & \cite{MartinezdeEscobar}\\
        $^{84}\textrm{Sr}$
            & Varying
                & 30
                     & 5
                        & 0.4
                            & 1.3
                                & 3.5
                                    & 0.1
                                        & 0.5
                                            & 0.02
                                                & 8.5
                                                    & 7.5
                                                        & \\

            & Fixed
                & 30
                     & 5
                        & 0.4
                            & 1.3
                                & 13
                                    & 1.1
                                        & 1.6
                                            & 2.6
                                                & 0.2
                                                    &
                                                        & \\
\hline
           & Experimental
                & 10
                    & 180
                        & 4.7
                            & \emph{n.a.}
                                & 0.05
                                    & 0.8
                                        & 4.7
                                            & \emph{n.a.}
                                                & 2
                                                    & 4
                                                        & \cite{PhysRevLett.91.040404}\\
        $^{174}\textrm{Yb}$
            & Varying
                & 10
                     & 180
                        & 4.7
                            & 0.6
                                &
                                    & \emph{No atoms left}
                                        &
                                            &
                                                &
                                                    & 4
                                                        & \\

            & Fixed
                & 10
                     & 180
                        & 4.7
                            & 0.6
                                & 0.08
                                    & 2.8
                                        & 20
                                            & 0.3
                                                & 0.2
                                                    & 4
                                                        & \\
\hline
\hline
\end {tabular}}
\caption{ Experimental data and numerical results for Bosons. The simulation stops when $D=1$ where as $N_f$ for the experimental data are given for a similar condition. \emph{n.a.} means: data not available.}\label{exp_vs_sim_Boson}
\end{center}
\end{table}

\begin{table}  
\begin{center}
{\normalsize\begin{tabular}[t]{@{}c c c@{}*{2}{@{}c@{}} @{}c c@{} *{2}{@{}c@{}} @{}c c c c}
\hline
\hline

            & Type
                & $N_i$
                     & $T_i$
                        & $n_i$
                            &$\Gamma_{el}^i$
                                & $N_f$
                                    & $T_f$
                                        & $n_f$
                                            & $\Gamma_{el}^f$
                                                & $t_{evap}$
                                                    & $\eta$
                                                        & Ref\\

            &
                & \footnotesize{($10^5$)}
                     & \footnotesize{($T_F$)}
                        & \footnotesize{($10^{14}\textrm{cm}^{-3}$)}
                            & \footnotesize{($10^3\textrm{s}^{-1}$)}
                                & \footnotesize{($10^5$)}
                                    & \footnotesize{($T_F$)}
                                        & \footnotesize{($10^{14}\textrm{cm}^{-3}$)}
                                            & \footnotesize{($10^3\textrm{s}^{-1}$)}
                                                & \footnotesize{(s)}
                                                    &
                                                        & \footnotesize{(Experimental)}\\
\hline

            & Experimental
                & 30
                     & 2.7
                        & 0.25
                            & 0.2
                                & 10
                                    & 1
                                        & \emph{n.a.}
                                            & \emph{n.a.}
                                                & 4
                                                    & 7
                                                        & \cite{desalvo2010degenerate}\\
        $^{87}\textrm{Sr}$
            & Varying
                & 30
                     & 2.7
                        & 0.25
                            & 0.15
                                & 3
                                    & 1
                                        & 0.03
                                            & 0.004
                                                & 36
                                                    & 7
                                                        & \\

            & Fixed
                & 30
                    & 2.7
                        & 0.25
                            & 0.15
                                & 10
                                    & 1
                                        & 0.8
                                            & 0.2
                                                & 1
                                                    & 7
                                                        & \\
\hline
           & Experimental
                & 3
                    & 100
                        & \emph{n.a.}
                            & \emph{n.a.}
                                & 0.013
                                    & 5.6
                                        & \emph{n.a.}
                                            & \emph{n.a.}
                                                & 2
                                                    & 7
                                                        & \cite{PhysRevLett.98.030401}\\
        $^{173}\textrm{Yb}$
            & Varying
                & 3
                     & 100
                        & 2.5
                            & 0.25
                                &
                                    & \emph{No atoms left}
                                        &
                                            &
                                                &
                                                    &
                                                        & \\

            & Fixed
                & 3
                     & 100
                        & 2.5
                            & 0.25
                                & 0.5
                                    & 6
                                        & 3
                                            & 0.8
                                                & 1
                                                    & 7
                                                        & \\
\hline
\hline
\end {tabular}}
\caption{ Experimental data and numerical results for Fermions. The simulation stops when $T/T_F=1$ where as $N_f$ for the experimental data are given for a similar condition. \emph{n.a.} means: data not available.} \label{exp_vs_sim_Fermion}
\end{center}
\end{table}

Initially, I set $\gamma_l(n,T)=0$ and three-body recombination losses will be discussed latter on. After simplification of equations (\ref{eq_T}) and (\ref{eq_N}), it turns out that $N$ and $T$ are linked though the simple relation:
\begin{equation}
\label{eq_NTRel}
T\propto N^\alpha.
\end{equation}
From equation (\ref{eq_NTRel}), one derives other relations among quantities of interest such as the peak spatial density $n_0$, the elastic collision rate $\gamma$, the phase space density $D$, or if it concerns Fermions, the temperature in Fermi unit $T/T_F$. These quantities also depend on the confinement of the gas and of the trap truncated energy $\epsilon_t$. Taking into account the gaussian shape of the dipole trap, expressions can be derived from \cite{PhysRevA.81.063620}. However when $\eta\gg 1$, the atoms are present at the bottom of the potential where it can be approximated by an harmonic potential characterized by its mean oscillation frequency $\omega$. In the fixed trap configuration, $\omega$ remains constant during the evaporation. In contrast, in the varying one where the total power of the trapping laser is reduced and with respect to the relations (\ref{eq_wP}), (\ref{eq_def_eta}) and (\ref{eq_NTRel}), one has $\omega\propto N^{\frac{\alpha}{2}}$. Both trap configurations can be modeled introducing a parameter $\theta$ such that
\begin{equation}
\label{eq_omegaN}
\omega\propto N^{\theta\frac{\alpha}{2}},
\end{equation}
where $\theta=0$($=1$) means that evaporation occurs in a fix (varying) 3D trap.
Finally, from relations (\ref{eq_NTRel})and (\ref{eq_omegaN}), one sets:
\begin{equation}
\label{eq_nN}
n_0\propto N^{1-\frac{3}{2}(1-\theta)\alpha},
\end{equation}
\begin{equation}
\label{eq_gN}
\gamma\propto N^{1-(1-\frac{3}{2}\theta)\alpha},
\end{equation}
\begin{equation}
\label{eq_DN}
D\propto N^{1-\frac{3}{2}(2-\theta)\alpha},
\end{equation}
\begin{equation}
\label{eq_TTF}
\frac{T}{T_F}\propto N^{-\frac{1}{3}+\frac{1}{2}(2-\theta)\alpha}.
\end{equation}
Since $\alpha>1$, these relations clearly show some differences among the two configurations ($\theta=0,1$). Firstly, the runaway regime, \emph{i.e.} an increase in $\gamma$, could be reached only if $\theta=0$. As a direct consequence, the cooling time is shortest for a fix trap. Secondly, and without any surprise, $D$ ($T/T_F$) increases (decreases) for both configurations. However, the $\theta=0$ configuration leads to a better efficiency of the evaporation.

One has to keep in mind however that relations (\ref{eq_NTRel}-\ref{eq_TTF}) are expressed only in a lossless trap ($\gamma_l(n,T)=0$) and for $\eta$ being large and constant. Those assumptions are not necessary fulfilled in some experiments. Thus, comparison between the model predictions and the documented evaporative schemes can be carried out through only with great care. Nevertheless an attempt is summarized in tables \ref{exp_vs_sim_Boson} $\&$ \ref{exp_vs_sim_Fermion} where the raw data of some successful experiments toward degenerate gases of two-electron atoms are compared with the results of the model presented in this work. As was already mentioned and discussed in the original papers \cite{PhysRevLett.103.200401}, \cite{MartinezdeEscobar} and also confirmed here, the experimental evaporation ramps for $^{84}\textrm{Sr}$ reasonably stick to the $\theta=1$ model. In contrast for $^{40}\textrm{Ca}$, $^{87}\textrm{Sr}$, $^{173}\textrm{Yb}$ and $^{174}\textrm{Yb}$, either of the two evaporation models matches with the experimental data. Here, the trap is made with two intensity independent dipole lasers to which more complex experimental ramps are done. It is worth mentioning however that experimental data might be constrained between the two trap configuration model, where the $\theta=0$ one stands to be the most efficient in term of evaporation time $t_{evap}$ and remaining atoms $N_f$.

\begin{figure}
\begin{center}
\includegraphics[scale=0.5]{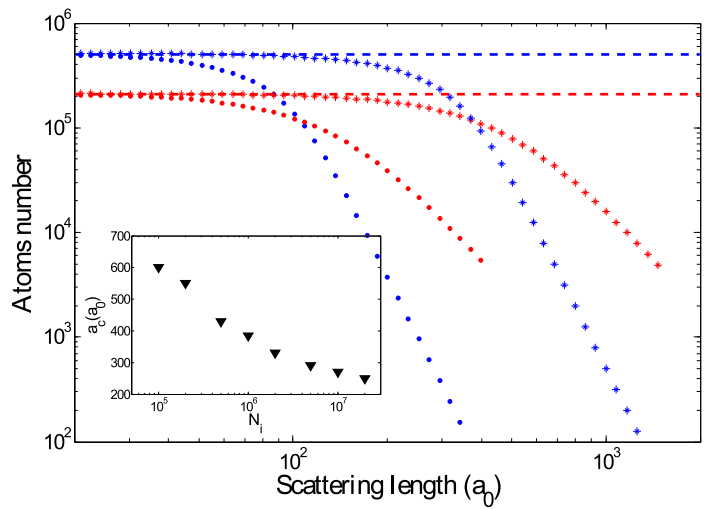}
\caption{Remaining atoms at $D_f=1$ after a forced evaporation for different scattering length values ranging form $10a_0$ to $2000a_0$. Each point correspond to a simulation at a given scattering length. For all points $\eta=10$, $N_i=10^6$, $T_i=20\mu K$ and $n_i=1.3\cdot 10^{14}\textrm{cm}^{-3}$ leading to an initial phase space density of $D_i=10^{-2}$. The blue (red) curves correspond to $\theta=0$ ($=1$) configuration where as the stars (dots) refereed to $n_lC=16.5 (210)$ (see text). The dash lines disregard the loss induced by the three-body recombination ($n_lC=0$) for the $\theta=0$ (blue) and $\theta=1$ (red) configuration. The inset shows, as function of the initial atoms number for $n_lC=16.5$, the critical values $a_c$ of the scattering length for which the two trap configurations give the same result.}
\label{3body}
\end{center}
\end{figure}

Lets us now analyze how the three-body recombination losses affect the previous results. Indeed, reference \cite{kraft2009bose} addresses three-body recombination as a major loss mechanism. It has been also reported that three-body recombination limit the elastic collision rate in a trap configuration close to the $\theta=0$ configuration \cite{PhysRevA.79.061406}. Firstly, I recall that the three-body recombination event scales with $n^3$. Thus equation (\ref{eq_nN}) and tables \ref{exp_vs_sim_Boson} $\&$ \ref{exp_vs_sim_Fermion} indicate that the fixed trap configuration may be more affected than the the varying one. The three-body recombination atoms loss rate is given by
\begin{equation}
\label{eq_3bALR}
\gamma_l(n,T)=L_3\langle n^2\rangle,
\end{equation}
with
\begin{equation}
\label{eq_L3}
L_3=n_lC\frac{\hbar}{m}a^4.
\end{equation}
Here, $n_l$ stands for the mean number of atoms lost per collision event. $0<C<70$ is a dimensionless factor that might also vary with the scattering length $a$ (see \cite{PhysRevLett.91.123201} and references therein). Both values of $n_l$ and $C$ are of crucial importance for quantitative evaluation of the three-body recombination losses, but unfortunately they are not well documented for two-electron atoms. In \cite{kraft2009bose}, the authors report $L_3=3\cdot 10^{-27}\textrm{cm}^6\textrm{/s}$ for $^{40}\textrm{Ca}$ which can be converted to $n_lC=16.5$. Plugging this value into the model, the efficiency of the evaporation is shown in figure \ref{3body}. The blue (red) stars correspond to the $\theta=0$ ($=1$) configuration. One notices that at low scattering length the three-body losses have a limited impact and the $\theta=0$ configuration remains the best option (for more details about the comparisons see the figure caption). However above a characteristic value of the scattering length, $a_c\sim 385a_0$ in that case, it seems more relevant to use the varying trap where the spatial density remains moderate. Similar conclusion can be drawn if one uses the most pessimistic value $n_lC=3\cdot 70$. In that case, the inversion of the efficiency among the two trap configuration is observed at lower value: $a_c\sim 105a_0$. The values of $a_c$ are non universal and may be figured out for any experimental realization. For example, in the inset of figure \ref{3body} is shown the variation of $a_c$ as function of $N_i$.

\section{Conclusion}\label{Sec_Conclusion}

If three-body recombination losses can be disregarded, the dynamic of the evaporation, at $\eta$ constant, is ruled by a set simple set of nonlinear equations for $N$ and $T$. The forced evaporation in two configurations of the dipole trap has been compared: the varying trap where the laser power of the dipole trap is ramped down leading to a reduction of the spatial confinement and the fix trap where the spacial confinement is kept constant. I show that the runaway regime, characterized by an increase of the elastic collision rate, is reached only in the fix trap configuration. As a consequence the efficiency of the evaporation in term of remaining atoms and cooling time is significantly improved. However, the spatial density increases during the evaporation and three-body recombination losses have also to be considered. In this context I show that there exists a characteristic scattering length above which the varying trap configuration, with lower spatial confinement, becomes the best choose.

A practical implementation of the fixed trap can be done using the $^3\!P_0$ long living excited state in two-electron atoms as an anti-trapping state. Just like with RF \emph{knife} for magnetic traps, the dipole trap can be truncated using an quasi resonant optical field on the $^1\!S_0\rightarrow\,^3\!P_0$ transition. The forced evaporation is obtained sweeping the laser detuning.

The discussion in this paper had been carried out considering the lower branch of the adiabatic potential (in blue figure \ref{87Sr}b). If the atoms are now transferred to the upper branch (in red figure \ref{87Sr}b), they are confined in a quasi 2D bubble-like optically-dressed trap. Similar types of traps have been already proposed \cite{PhysRevLett.86.1195} and accomplished \cite{0295-5075-67-4-593} with RF-dressed state in magnetic traps. In those, the interplay between the Zeeman sublevels structure and the vectorial nature of the RF field creates some holes through which evaporation is done \cite{0953-4075-40-20-004}. In a bubble-like optically-dressed trap, such holes do not exist at least in the two levels scheme considered here. However, forced evaporation can still take place sweeping the dressed laser frequency in the opposite direction to the one used for the forced evaporation in the lower branch of the adiabatic potential.

\section{Acknowledgments}

My acknowledgments go to Wenhui Li for fruitful comments on the manuscript. I thank the Section 4, INP and DRI institutions of CNRS for financial supports during my stay in Singapore.

\section{Bibliography}

\bibliographystyle{unsrt}	
\bibliography{Quantum_Gas,Alkaline_earth}

\end{document}